# Does TMS increase BOLD activity at the site of stimulation?

Farshad Rafiei & Dobromir Rahnev

School of Psychology, Georgia Institute of Technology, Atlanta, GA

**Keywords**: concurrent TMS-fMRI, BOLD, single-neuron recording

**Acknowledgments**

This work was supported by the National Institute of Health (awards: R01MH119189 & R21MH122825).

**Competing interests:**

None

**Correspondence**

Farshad Rafiei

Georgia Institute of Technology

654 Cherry Str. NW

Atlanta, GA 30332

E-mail: farshad@gatech.edu

**Abstract**

Transcranial magnetic stimulation (TMS) is widely used for understanding brain function in neurologically intact subjects and for the treatment of various disorders. However, the precise neurophysiological effects of TMS at the site of stimulation remain poorly understood. The local effects of TMS can be studied using concurrent TMS-fMRI, a technique where TMS is delivered during fMRI scanning. However, although concurrent TMS-fMRI was developed over 20 years ago and dozens of studies have used this technique, there is still no consensus on whether TMS increases blood-oxygen-level-dependent (BOLD) activity at the site of stimulation. To address this question, here we review all previous concurrent TMS-fMRI studies that reported analyses of BOLD activity at the target location. We find evidence that TMS increases local BOLD activity when stimulating the primary motor and visual cortices but that these effects are likely driven by the downstream consequences of TMS (finger twitches and phosphenes). However, TMS does not appear to increase BOLD activity at the site of stimulation for areas outside of the primary motor and visual cortices when conducted at rest. We examine the possible reasons for such lack of BOLD signal increase based on recent work in non-human animals. We argue that the current evidence points to TMS inducing periods of increased and decreased neuronal firing that mostly cancel each other out and therefore lead to no change in the overall BOLD signal.

## Introduction

Transcranial magnetic stimulation (TMS) is a non-invasive technique commonly used in both basic science research and clinical interventions (Chail, Saini, Bhat, Srivastava, & Chauhan, 2018; Kobayashi & Pascual-Leone, 2003; Ziemann, 2017). TMS induces electric currents in the brain via rapidly alternating magnetic fields (Sack & Linden, 2003). These induced electric currents can affect the ongoing neural activity patterns by depolarizing membrane potentials in local cortical brain tissue at the site of stimulation (Reithler, Peters, & Sack, 2011) and can transiently interfere with the pattern of activity in a neural population which is involved in processing an ongoing task (Chambers & Mattingley, 2005; Ruff, Driver, & Bestmann, 2009; Walsh & Cowey, 2000). Different TMS protocols can either impair (Bestmann, Ruff, et al., 2008; Ruff et al., 2009) or enhance (Reis et al., 2008; Reithler et al., 2011) task performance, making TMS a flexible technique well suited to both clinical applications and a variety of basic science questions.

The importance of TMS in both basic and clinical research necessitates that one understands well the neurophysiological effects of TMS at the site of stimulation. To explore this question, TMS has been combined with several neuroimaging techniques such as electroencephalography (EEG) (Oliviero, Strens, Di Lazzaro, Tonali, & Brown, 2003; Paus, Sipila, & Strafella, 2001; Rossi, 2000; Strens et al., 2002; Thut et al., 2003), positron emission tomography (PET) (Chouinard, Van Der Werf, Leonard, & Paus, 2003; Ferrarelli et al., 2004; Fox et al., 1997; Knoch et al., 2006; Paus et al., 2001, 1997, 1998; Siebner et al., 2001) and single photon emission computed tomography (SPECT) (Okabe et al., 2003). A drawback for all of these methods, however, is the relatively lower spatial resolution, which makes it difficult to resolve the activity induced specifically at the site of stimulation. Therefore, it has become increasingly popular to combine TMS with functional magnetic resonance imaging (fMRI) in order to examine how the stimulation affects the blood-oxygen-level-dependent (BOLD) signal immediately underneath the TMS coil (Bestmann, Ruff, et al., 2008; Driver, Blankenburg, Bestmann, Vanduffel, & Ruff, 2009; Ruff et al., 2009).

The first study to employ concurrent TMS-fMRI was published in 1997 and focused on demonstrating the feasibility of the technique by examining the 3D intensity maps of the TMS-induced magnetic field inside a conventional MR scanner (Bohning et al., 1997). Since then, many studies have used concurrent TMS-fMRI to reveal the local and distant effects of TMS during rest or task execution. The field has witnessed substantial improvements in the development of MRI-compatible TMS equipment including TMS coils, neuronavigation, stands, and even special MRI receiver coils designed for use in concurrent TMS-fMRI studies (De Weijer et al., 2014; Navarro De Lara et al., 2015; Wang, Xu, & Butman, 2017). Such improvements have made it possible to investigate the BOLD activity at the precise site of stimulation with relatively high resolution and with negligible signal loss compared to standard fMRI.

However, despite all of these technical improvements and the availability of extensive research, the basic question of whether and under what conditions TMS affects the BOLD signal at the site of stimulation remains unresolved. At a first glance, the dozens of studies that have addressed this question have found conflicting results. To complicate matters, these studies differ in many aspects such as the intensity and duration of stimulation, the stimulation site, the types of analyses performed, the precision of identifying the site of stimulation, etc. This variability between studies has made it challenging to arrive at a unified understanding of the effects of TMS on local BOLD activity. Instead, the fact that TMS induces neuronal firing under the coil, which has been established beyond doubt in studies of non-human animals (Aydin-Abidin, Moliadze, Eysel, & Funke, 2006; Moliadze, Giannikopoulos, Eysel, & Funke, 2005; Moliadze, Zhao, Eysel, & Funke, 2003; Pasley, Allen, & Freeman, 2009), has led to the default assumption that TMS must also increase BOLD activity at the site of stimulation.

The goal of this paper is to assemble and critically examine the currently available evidence regarding the BOLD effects of TMS at the site of stimulation. We start by considering the factors that are most critical to establishing the local TMS effects on BOLD activity. We then systematically review all published papers on concurrent TMS-fMRI that reported any analyses of the BOLD signal at the site of stimulation and specifically examine how they relate to the factors we have identified. The picture that emerges is that TMS increases BOLD activity at the site of stimulation when it results in downstream effects such as hand twitches of phosphenes, but that it does not affect local BOLD activity otherwise. Finally, we explore how these findings relate to the available animal research on the effects of TMS on the firing rate of single neurons.

**Factors critical to establishing the TMS effects on local BOLD activity**

Establishing the BOLD effects of TMS at the site of stimulation requires the consideration of several factors. Here we identify four factors (**Table 1**) that we believe are most critical to understanding the local effects of TMS and examine where each published study on concurrent TMS-fMRI stands in relation to these factors. The first three factors relate to what aspects of stimulation determine the effect on local BOLD activity; the last is about the methodological considerations related to the precision of TMS localization and type of analysis.

**Table 1. Four critical factors**. The table lists factors likely to be critical for establishing the effect of TMS at the site of stimulation and reasons for the importance of each factor. The first three factors relate to aspects of stimulation likely to affect the BOLD signal, whereas the last factor is methodological.

| Factor | Why is this factor important? |
|---|---|
| Site of stimulation | Downstream effects of M1 or V1 stimulation may lead to different results for these sites compared to others |
| Task vs. rest | TMS may have different effects based on whether a subject is at rest or engages in a task |
| Intensity and amount of stimulation | Higher intensities and more pulses may be more likely to affect the BOLD activity |
| Precision of TMS localization and type of analyses | Imprecise localization of the site of stimulation increases the possibility for both false positives and false negatives; analyses on precisely localized ROIs are likely to be most informative |

Site of stimulation

The local effects of TMS are likely not identical throughout the brain. It is well known that different areas of the brain have different cytoarchitecture (Fischl et al., 2008; Roth, Padberg, & Zangen, 2007), which may contribute to differential effects on local BOLD activity. It is also possible that the exact orientation of the brain tissue at the site of stimulation affects the observed BOLD. Unfortunately, the published literature is mostly insufficient to address the effects of these factors since information is rarely available about which Brodmann area was stimulated (and whether this was consistent across all subjects) or the orientation of the brain tissue underneath the TMS coil.

However, it is still possible to make a more coarse-grained distinction between studies that targeted the primary motor cortex (M1) or the primary visual cortex (V1) versus studies that targeted the rest of the brain. The reason for this categorization is that both M1 and V1 TMS can lead to downstream consequences that a subject can directly experience, unlike targeting the rest of the brain where the only subjective experience produced has to do with the auditory and tactile consequences of TMS. Specifically, targeting M1 can lead to twitches in the contralateral hand (or other body parts), while targeting V1 can lead to subjective visual experiences called phosphenes. To address this distinction, we consider the studies that targeted M1/V1 in a separate section from all remaining studies.

### Rest vs. task

The majority of concurrent TMS-fMRI studies are conducted at rest with the idea to investigate the effects of neural signals specifically created by TMS. However, an increasing proportion of studies are now conducted with an accompanying task to test how TMS modulates brain activity and connectivity. Critically, the effects of TMS on local BOLD activity may depend on whether the targeted region is already engaged in a task. To address this issue, we further split the studies targeting areas outside of M1/V1 into studies conducted at rest vs. during task (all published concurrent TMS-fMRI studies targeting M1/V1 were done at rest).

### The strength and amount of stimulation

Higher TMS intensities create larger magnetic fields, which in turn create more neural firing (Aydin-Abidin et al., 2006; Bergmann, Karabanov, Hartwigsen, Thielscher, & Siebner, 2016; Fitzgerald, Fountain, & Daskalakis, 2006; Krieg, Salinas, Narayana, Fox, & Mogul, 2015; Matheson, Shemmell, De Ridder, & Reynolds, 2016). Similarly, a larger amount of stimulation in a given period (i.e., more TMS pulses) is also likely to drive up neural firing. Other features of TMS delivery such as the frequency or pattern of stimulation may also be critical. Therefore, we explicitly consider the intensity of stimulation, the number of pulses delivered in a single stimulation period, and the length of the stimulation period for all studies we discuss.

Precision of TMS localization and type of analyses

Perhaps the most important methodological issue to consider when examining the effects of TMS on local BOLD activity is the precision with which the site of stimulation has been localized. Precise localization allows the researcher to test only the site immediately underneath the coil with high power (e.g., without the need to correct for multiple comparisons). On the other hand, imprecise or non-existent localization necessitates that a larger area is examined, which increases the likelihood of both false positives (e.g., an activation is found that is not at the true site of stimulation) and false negatives (e.g., the need to correct for multiple comparisons obscures a significant effect at the site of stimulation). In addition, TMS targeting inside the scanner is always subject to error. Therefore, second-level analyses – which involve conducting across-subject analyses on normalized individual subjects' brains – are likely to be particularly noisy since different subjects end up receiving stimulation at different sites (in normalized space) with the variability often being substantial (Vink et al., 2018). Therefore, we treat the studies that precisely localized the site of stimulation for each subject and conducted region-of-interest (ROI) analyses as the gold standard for revealing the TMS effects on local BOLD activity.

Critically, the importance of the precision of TMS localization has been confirmed empirically. A study that recorded neuronal firing in the vicinity of the TMS coil in awake monkeys found that TMS had little to no effect on neurons located as close as 2 mm away from the center of the TMS coil (Romero, Davare, Armendariz, & Janssen, 2019). If the focality of these TMS effects is similar in the human brain, then the results from studies that do not precisely localize the actual location of the TMS coil for each subject likely do not reflect the effects of TMS at the exact site of stimulation.

**Studies that targeted M1 or V1**

M1 and V1 are unique among typical TMS target locations in that stimulation to those areas produces effects that the participant has a subjective experience of. Specifically, TMS to M1 leads to motor twitches and TMS to V1 leads to visual experiences called "phosphenes." Most studies targeting M1 delivered TMS as a proportion of the resting motor threshold (rMT), which is the stimulation intensity for which a single TMS pulse over M1 elicits a contralateral hand twitch on half of the trials when the target muscle is at rest. Therefore, studies that stimulate at 100% or higher of rMT elicit hand twitches, and consequently these stimulation intensities are referred to as "supra-threshold," with lower intensities referred to as "sub-threshold." We also note that several studies have used intensities expressed as a proportion of active motor threshold (aMT), which is the stimulation intensity for which a single TMS pulse over M1 elicits a contralateral hand twitch on half of the trials during active contraction of the target muscle. The aMT is typically lower than the rMT, and therefore even 110% of aMT is usually not enough to elicit muscle twitches at rest and is therefore classified as sub-threshold stimulation. Similar to the rMT, the single study targeting V1 delivered TMS as a proportion of phosphene threshold (PT) with intensities of 100% of PT leading to the subjective experience of phosphenes on approximately half the trials.

<u>Supra-threshold TMS induces a local BOLD increase</u>

To date, 15 concurrent TMS-fMRI studies have targeted M1 or V1 with supra-threshold TMS intensities during rest (Baudewig et al., 2001; Bestmann, Baudewig, Siebner, Rothwell, & Frahm, 2003, 2004; Bohning et al., 1999; Bohning, Shastri, McGavin, et al., 2000; Bohning et al., 1998; Bohning, Shastri, Wassermann, et al., 2000; Caparelli et al., 2010; De Weijer et al., 2014; Denslow, Bohning, Bohning, Lomarev, & George, 2005; Denslow, Lomarev, George, & Bohning, 2005; Jung, Bungert, Bowtell, & Jackson, 2016, 2020; Kemna & Gembris, 2003; Shitara, Shinozaki, Takagishi, Honda, & Hanakawa, 2011). Fourteen of these studies reported significant BOLD activations at or near the site of stimulation (**Table 2**) with only one study failing to find such activations (Jung et al., 2020). Fourteen of these studies targeted M1, while only one of them targeted V1 (Caparelli et al., 2010).

**Table 2. Studies targeting M1/V1**. The table first lists all 15 studies using supra-threshold stimulation first in order of publication. All five studies that reported sub-threshold contrasts (reported in the second part of the table) also appear in the first part of the table. aMT, active motor threshold; GLM, general linear model; HRF, hemodynamic response function; L, left; rMT, resting motor threshold; PT, phosphene threshold; ROI, region of interest.

| Study | Target | Protocol | Contrast(s) | N | Activation | Analyses |
|---|---|---|---|---|---|---|
| Bohning et al. (1998) *Invest Radiol* | L M1 | 20 pulses over 24 s | 110% rMT > rest | 3 | Yes | First-level GLMs |
| Bohning et al. (1999) *Biol Psychiatry* | L M1 | 18 pulses over 17 s | 110% rMT > rest | 7 | Yes | 1) ROI (actual coil position)<br>2) HRF time-courses in ROI |
| Bohning et al. (2000) *J Magn Reson Imaging* | L M1 | single pulse | 120% rMT > rest | 5 | Yes | 1) ROI (actual coil position)<br>2) HRF time-courses in ROI |
| Bohning et al. (2000) *Invest Radiol* | L M1 | 21 pulses over 20 s | 110% rMT > rest | 7 | Yes | Unknown |
| Baudewig et al. (2001) *NeuroReport* | L M1 | 10 pulses over 1 s | 110% rMT > zero | 6 | Yes | HRF time-courses in ROI defined from a different task |
| Bestmann et al. (2003) *NeuroImage* | L M1 | 40 pulses over 10 s | 110% rMT > zero | 8 | Yes | 1) ROI (anatomically-defined)<br>2) HRF time-courses in ROI |
| Kemna and Gembris (2003) *Neurosci Lett* | L M1 | 4 pulses over 1 s | 150% rMT > zero | 8 | Yes | 1) ROI (actual coil position)<br>2) HRF time-courses in ROI |
| Bestmann et al. (2004) *Eur J Neurosci* | L M1 | 30 pulses over 9.6 s | 110% rMT > 90% aMT<br>110% rMT > rest | 11 | Yes | 1) ROI (anatomically-defined)<br>2) Second-level GLM |
| Denslow et al. (2005) *Biol Psychiatry* | L M1 | 21 pulses over 21 s | 110% rMT > rest | 11 | Yes | 1) ROI (anatomically-defined)<br>2) Second-level GLM |
| Denslow et al. (2005) *Cogn Behav Neurol* | L M1 | 21 pulses over 21 s | 110% rMT > rest | 9 | Yes | 1) ROI (anatomically- and functionally-defined) 2) HRF time-courses in ROI |
| Caparelli et al. (2010) *Open Neuroimag J* | L V1 | 8 pulses over 28 s | 100% PT > rest | 12 | Yes | Second-level GLM |
| Shitara et al. (2011) *NeuroImage* | L M1 | single pulse | 120% rMT > rest | 36 | Yes | 1) ROI (anatomically-defined) 2) Second-level GLM |
| De Weijer (2014) *J Clin Neurophysiol* | L M1 | single pulse | 110% rMT > 70% rMT | 4 | Yes | 1) ROI (actual coil position) 2) First-level GLMs |
| Jung et al. (2016) *Brain Stimul* | L M1 | 12 pulses over 12 s | 100% rMT > vertex TMS | 32 | Yes | Second-level GLM |
| Jung et al. (2020) *Front Hum Neurosci* | L M1 | 11 pulses over 11 s | 100% rMT > rest | 12 | No | Second-level GLM |
| Bohning et al. (1999) *Biol Psychiatry* | L M1 | 18 pulses over 17 s | 80% rMT > rest | 7 | No | 1) ROI (actual coil position) 2) HRF time-courses in ROI |
| Baudewig et al. (2001) *NeuroReport* | L M1 | 10 pulses over 1 s | 90% rMT > zero | 6 | No | HRF time-courses in ROI defined from a different task |
| Bestmann et al. (2003) *NeuroImage* | L M1 | 40 pulses over 10 s | 110% aMT > zero 90% aMT > zero | 8 | No | 1) ROI (anatomically-defined) 2) HRF time-courses in ROI |
| Bestmann et al. (2004) *Eur J Neurosci* | L M1 | 30 pulses over 9.6 s | 90% aMT > rest | 11 | No | 1) ROI (anatomically-defined) 2) Second-level GLM |
| Shitara et al. (2011) *NeuroImage* | L M1 | single pulse | 120% rMT > 90% aMT | 36 | No | 1) ROI (anatomically-defined) 2) Second-level GLM |

In one example study, which is also the very first concurrent TMS-fMRI study that reported brain results, Bohning and colleagues applied single TMS pulses at 110% of rMT over M1. They then conducted a general linear model (GLM) analysis for each subject separately (i.e., first-level analysis) comparing the TMS condition to rest, and found a significant BOLD signal increase near the site of stimulation (Bohning et al., 1998). Another study (Jung et al., 2016) performed similar analyses but at the group (second) level and also found significant activations in the vicinity of the site of stimulation. These types of analyses, however, do not ensure that the activations are at the precise site of stimulation.

Therefore, most subsequent research defined the site of stimulation as an ROI. In six studies, the ROI was defined based on either anatomical considerations or functional activations from a related task (Baudewig et al., 2001; Bestmann et al., 2003, 2004; Denslow, Bohning, et al., 2005; Denslow, Lomarev, et al., 2005; Shitara et al., 2011). Although this method is superior to either first- or second-level analyses, these ROIs still lack precision and may not reflect the exact TMS coil location. Critically, four studies defined an ROI at the precise site of stimulation using markers placed directly on the TMS coil (Bohning et al., 1999; Bohning, Shastri, Wassermann, et al., 2000; De Weijer et al., 2014; Kemna & Gembris, 2003). All 10 of these studies reported significant activity at the site of stimulation, though in one of the studies the observed activations were slightly shifted from the actual coil position (De Weijer et al., 2014). Overall, these studies paint a picture where TMS to M1 consistently increases BOLD activity at the site of stimulation.

Only one study to date has failed to observe local BOLD increase after TMS to M1 (Jung et al., 2020). In that study, three groups of subjects were compared with one of them receiving TMS to M1 at rest. Although this was not the focus of the study, the authors conducted a second-level analysis of TMS > rest for that group (N = 12) and did not report any activations in the motor cortex. However, given the conservativeness of second-level analysis for finding activations in a pre-specified region and the relatively small sample size for such second-level analyses, this null result should be interpreted with caution.

To date, a single concurrent TMS-fMRI study has targeted V1 (Caparelli et al., 2010). The authors stimulated at 100% of the phosphene threshold, yet some subjects experienced phosphenes while others did not. The results showed that there was a significant BOLD increase in the visual cortex for the group that experienced phosphenes but not for the group that did not experience phosphenes. However, these results are based on second-level analyses and come from a single study. Nonetheless, it can be provisionally hypothesized that V1 TMS may act similarly to M1 TMS where local BOLD activity occurs only for supra-threshold stimulation that has observable downstream effects.

Sub-threshold TMS does not increase the local BOLD signal

Five of the studies above included control conditions with sub-threshold TMS intensities (i.e., below 100% of rMT). All five of the studies found that such sub-threshold stimulation did not affect the overall BOLD activity at the site of stimulation (Baudewig et al., 2001; Bestmann et al., 2003, 2004; Bohning et al., 1999; Shitara et al., 2011). In all of these cases, the null results were found in the same ROIs where a significant BOLD increase was found for supra-threshold stimulation. These null results were reported for 80% of rMT (Bohning et al., 1999), 90% of rMT (Baudewig et al., 2001), 90% of aMT (Bestmann et al., 2003, 2004; Shitara et al., 2011), and 110% of aMT (Bestmann et al., 2003), with the last intensity still being sub-threshold because it is lower than 100% of rMT. These results suggest that even intensities slightly lower than 100% of rMT do not induce BOLD signal increases at the site of stimulation in M1.

Summary

Overall, the studies that targeted M1 and V1 paint a very consistent picture: supra-threshold TMS intensities almost invariably lead to increases in local BOLD activity, whereas sub-threshold intensities invariably lead to no significant local BOLD increases. While it is possible that this pattern of results is due to the direct influence of TMS on the stimulated neural tissue, another plausible explanation is that the supra-threshold activations are a direct consequence of the downstream effects of supra-threshold stimulation. Indeed, many researchers have speculated that, in the case of M1 TMS, the observed activations are likely due to the contribution of afferent feedback from contralateral muscle responses (Bestmann & Feredoes, 2013; Bestmann, Ruff, et al., 2008). A related way in which supra-threshold M1 stimulation can induce local activity is if subjects, after detecting their hand twitch, consciously or unconsciously send motor commands to stabilize the hand and these putative post-TMS motor commands also contribute to the observed M1 BOLD increase.

Similarly, consciously perceived phosphenes after supra-threshold V1 stimulation can lead to higher-order areas sending feedback signals to the primary visual cortex. This mechanism is especially likely for V1 TMS since the study that targeted V1 (Caparelli et al., 2010) reported increased BOLD activity only for those subjects who perceived phosphenes even though the rest of the subjects experienced similar TMS intensities.

Overall, it appears likely that supra-threshold stimulation leads to local BOLD increases because of the downstream consequences it produces rather than because of the direct neural effects on the underlying brain tissue. Therefore, to convincingly establish the direct neural effects of TMS on local BOLD activity, it is important to examine the results of studies that target areas outside of the primary motor and sensory cortices.

**Studies that targeted the rest of the brain during rest**

As highlighted already, different TMS-related effects on local BOLD activity can be expected during rest vs. tasks that require the engagement of the targeted region (**Table 1**). Therefore, for the studies targeting areas outside of M1/V1, we separately consider the ones conducted at rest vs. during task. This section focuses on the studies at rest.

Studies reporting an increase in BOLD activity

A total of 15 concurrent TMS-fMRI studies that targeted areas outside M1/V1 during rest have been published (Baudewig et al., 2001; Bestmann, Baudewig, Siebner, Rothwell, & Frahm, 2005; Blankenburg et al., 2008; de Vries et al., 2009; De Weijer et al., 2014; Dowdle, Brown, George, & Hanlon, 2018; Jung et al., 2016; Kearney-Ramos et al., 2018; Kemna & Gembris, 2003; Leitão, Thielscher, Werner, Pohmann, & Noppeney, 2013; X. Li, Nahas, et al., 2004; X. Li, Tenebäek, et al., 2004; Rafiei, Safrin, Wokke, Lau, & Rahnev, 2021; Sack et al., 2007; Vink et al., 2018). Of these, three studies reported significant BOLD increases at or near the site of stimulation, while the remaining 12 reported no local change in BOLD. Here we discuss in detail the three studies that found increased BOLD; the next subsection discusses the remaining 12 studies.

**Table 3. Studies delivering TMS outside of M1/V1 during rest**. The table lists all 16 individual experiments using concurrent TMS-fMRI and stimulating areas outside M1 or V1 during rest. The first three experiments reported significant activations in the vicinity of the TMS coil, whereas the remaining 13 reported no significant activations. aMT, active motor threshold; rMT, resting motor threshold; GLM, general linear model; HRF, hemodynamic response function; L, left; MSO, maximum stimulator output; PFC, prefrontal cortex; PMd, dorsal premotor cortex; DLPFC, dorsolateral prefrontal cortex; PMC, premotor cortex; SPL, superior parietal lobule; IPS, intraparietal sulcus; dSMG, dorsal supramarginal gyrus; VMPFC, ventromedial prefrontal cortex; R, right; L, left; ROI, region of interest.

| Study | Target(s) | Protocol(s) | Contrast(s) | N | Activation | Analyses |
|---|---|---|---|---|---|---|
| Li et al. (2004) *Biol Psychiatry* | L PFC | 21 pulses over 21 s | 100% rMT > rest | 14 | Yes | Second-level GLM |
| Bestmann et al. (2005) *NeuroImage* | L PMd | 30 pulses over 10 s | 1) 110% rMT > 90% aMT<br>2) 110% rMT > rest | 9 | Yes | 1) ROI (anatomically-defined)<br>2) Second-level GLM |
| Vink et al. (2018) *Hum Brain Mapp* | L DLPFC | single pulse | 115% rMT > 60% rMT | 9 | Yes | First-level GLMs |
| Baudewig et al. (2001) *NeuroReport* | L PMC | 10 pulses over 1 s | 110% rMT > zero | 6 | No | HRF time-courses in ROI defined from a different task |
| Kemna and Gembris (2003) *Neurosci Lett* | 1) L PFC<br>2) L parietal | 4 pulses over 1 s | 150% rMT > zero | 8 | No | 1) ROI (actual coil position)<br>2) HRF time-courses in ROI |
| Li et al. (2004) *NPP* | L PFC | 21 pulses over 21 s | 1) 120% rMT > rest<br>2) 100% rMT > rest<br>3) 120% rMT > 100% | 8 | No | Second-level GLM |
| Sack et al. (2007) *Cereb Cortex* | 1) L SPL<br>2) R SPL | 5 pulses over 300 ms | 100% MSO > rest | 8 | No | Second-level GLM |
| Blankenburg et al. (2008) *J Neurosci* | R parietal | 5 pulses over 500 ms | 110% rMT > 50% rMT | 5 | No | Second-level GLM |
| De Vries et al. (2009) *Brain Res* | L SPL | 10 pulses over 10 s | 115% rMT > rest | 10 | No | Second-level GLM |
| Leitão et al. (2013) *Cereb Cortex* | R IPS | 38 pulses over 20 s | 1) 66% MSO > 33% MSO<br>2) 66% MSO > rest | 20 | No | Second-level GLM |
| De Weijer (2014) *J Clin Neurophysiol* | L dSMG | single pulse | 110% rMT > 70% rMT | 3 | No | 1) ROI (actual coil position)<br>2) First-level GLMs |
| Jung et al. (2016) *Brain Stimul* | Vertex | 12 pulses over 12 s | 100% rMT > rest | 32 | No | Second-level GLM |
| Dowdle et al. (2018) *Brain Stimul* | L DLPFC | single pulse | 90-120% rMT > sham | 20 | No | 1) ROI (anatomically-defined)<br>2) Second-level GLM |
| Kearney-Ramos et al. (2018) *Sci Rep* | R VMPFC | single pulse | 100% rMT > rest | 49 | No | 1) ROI (anatomically-defined)<br>2) Second-level GLM |
| Rafiei et al. (2021) *Hum Brain Mapp, Expt 1* | R PFC | 1) 20 pulses over 10 s<br>2) 10 pulses over 10 s | 100% rMT > 50% rMT | 5 | No | 1) ROI (actual coil position)<br>2) First-level GLMs |
| Rafiei et al. (2021) *Hum Brain Mapp, Expt 2* | L DLPFC | 1) 30 pulses over 1.2 s<br>2) 30 pulses over 2.4 s<br>3) 30 pulses over 3.6 s<br>4) 30 pulses over 6 s | 100% rMT > rest | 6 | No | 1) ROI (actual coil position)<br>2) First-level GLMs |

The first study targeted left DLPFC with single TMS pulses of either 115% or 60% of rMT (Vink et al., 2018). The authors then compared these two conditions and found that the majority of individual subjects showed activation somewhere in the left prefrontal cortex. However, the left prefrontal cortex is a large area and therefore the observed activity may not be at the actual site of stimulation. Unfortunately, even though the authors were able to precisely localize the location of the TMS coil for each subject, they did not use this information to specifically examine the BOLD activity at the site of stimulation. Given that the supra-threshold TMS is louder and elicits more pronounced tactile sensations, it may have an indirect effect on many regions associated with cognitive and emotional responses. Therefore, in the absence of information regarding the activity at the precise site of stimulation, activations in the left prefrontal cortex (which is involved in both cognition and emotion) do not provide strong evidence regarding the direct local effects of TMS on BOLD activity.

The second study targeted the left prefrontal cortex (PFC) of depressed patients and delivered 1 Hz TMS for 21 seconds at 100% of rMT (X. Li, Nahas, et al., 2004). The authors normalized each subject's brain and reported second-level analyses of the whole group. These analyses revealed several activations throughout the brain with one prominent hot spot in the left PFC near the presumed site of stimulation. Importantly, individual-level information about the precise site of stimulation was not available as the authors targeted a region that was 5 cm rostral of M1 (following standard practice for localizing the site for stimulation in treatments for depression). This type of targeting leads to substantial variability in stimulation locations across subjects (Vink et al., 2018), which means that activation in a second-level map is likely to be away from the site of stimulation for a substantial proportion of individual subjects. Further, it is again possible that some of the reported activations in the study were due to non-specific effects of TMS. Finally, the same authors used an identical TMS protocol in a different study of healthy subjects and observed no significant activations in left PFC for both 100% and 120% of rMT (X. Li, Tenebäek, et al., 2004). Together, these considerations and additional results raise doubts that the activations reported by X. Li, Nahas, et al. (2004) were due to direct neural effects of TMS at the precise site of stimulation.

The final study that reported increased BOLD at or near the site of stimulation targeted the dorsal premotor cortex (PMd) using 3 Hz TMS delivered for 10 seconds (Bestmann et al., 2005). The authors employed both supra-threshold (110% of rMT) and sub-threshold (90% of aMT) stimulation and conducted two different sets of analyses. First, they performed a second-level analysis that revealed activation in the vicinity of the targeted area for supra-threshold compared to sub-threshold stimulation. However, similar to the study discussed in the previous paragraph (X. Li, Nahas, et al., 2004), coil position was determined relative to the motor hot spot that gives rise to phosphenes (the final position was 2 cm anterior and 1 cm medial to the hot spot), a procedure that is likely to lead to substantial variability in the stimulation location across subjects making it hard to know how close the observed activation was to the actual stimulation site. Second, the authors also conducted an ROI-based analysis, in which the PMd was defined for each subject separately based on a priori anatomical criteria developed in a previous paper (Bestmann et al., 2004). The authors found that supra-threshold, but not sub-threshold, stimulation evoked significant response in the PMd ROI but no statistical test on the comparison between the two types of stimulation was reported. Importantly, the second-level analysis uncovered widespread activations in motor-related areas of the brain including bilateral dorsal and ventral premotor cortex, supplementary motor area, cingulate gyrus, and right cerebellum. In fact, the authors emphasized "the close spatial correspondence between brain regions activated during voluntary finger movements and rTMS of the left PMd" (p. 24). These results demonstrate that TMS over PMd resulted in generalized response in motor-related areas, thus casting further doubt on whether the activation at the site of stimulation is the result of the direct neural effects of TMS.

Overall, none of the three studies to date that delivered TMS outside of M1/V1 and reported significant BOLD activations at the site of stimulation provide strong evidence that these activations are due to the direct neural effects of TMS. The first two studies provide at best weak evidence for this possibility given that neither localized the area of stimulation and both interpreted first-level or second-level activations in the vicinity of the assumed coil position as being directly underneath the coil (X. Li, Nahas, et al., 2004; Vink et al., 2018). On the other hand, the third study (Bestmann et al., 2005) provides only slightly stronger evidence for a direct neural effect of TMS given the non-specificity of the activation result, the lack of direct comparison in the ROI analysis between supra- and sub-threshold stimulation, and the fact that the ROIs were defined anatomically rather than based on the actual location of stimulation.

Studies reporting no increase in BOLD activity

Apart from the three studies discussed above that reported BOLD increases in the vicinity of the site of stimulation, 12 studies including a total of 13 individual experiments reported no such increases (Baudewig et al., 2001; Blankenburg et al., 2008; de Vries et al., 2009; De Weijer et al., 2014; Dowdle et al., 2018; Jung et al., 2016; Kearney-Ramos et al., 2018; Kemna & Gembris, 2003; Leitão et al., 2013; X. Li, Tenebäek, et al., 2004; Rafiei et al., 2021; Sack et al., 2007). These experiments differed in several different dimensions. For example, four stimulated the right hemisphere, six stimulated the left hemisphere, one stimulated both hemispheres in different conditions, and one stimulated the vertex (which lies at the midline between the hemispheres). In addition, six stimulated the frontal lobe, five stimulated the parietal lobe, and one stimulated both the frontal and the parietal lobes in different conditions. The studies also varied in the number of pulses delivered at a time from a single TMS pulse (Dowdle et al., 2018; Kearney-Ramos et al., 2018) to 38 pulses over 20 seconds (Leitão et al., 2013). Further, all of these studies delivered TMS at intensities equal to or higher than 100% of rMT (two studies, Sack et al., 2007, and Leitão et al., 2013, chose intensities as a function of maximum stimulator output with the intensities corresponding to 120% and 126% of the average rMT). The diversity among these studies makes it less likely that any particular study feature can account for the lack of activation at the site of stimulation. Below we give more detail on these 12 studies with a focus on the precision of localizing the site of stimulation.

The results in six of the 12 studies were based on second-level GLMs (Blankenburg et al., 2008; de Vries et al., 2009; Jung et al., 2016; Leitão et al., 2013; X. Li, Tenebäek, et al., 2004; Sack et al., 2007). For example, one of these studies used five pulses at 10 Hz to stimulate the right parietal cortex and observed no activations near the stimulation spot when comparing supra-threshold (110% of rMT) and sub-threshold (50% of rMT) stimulation (Blankenburg et al., 2008). It is, however, important not to overinterpret the null results of such analyses. As already mentioned, the precise site of stimulation is certain to vary substantially from person to person, and therefore, even if TMS leads to local BOLD increases, these effects may not be visible in a second-level analysis. In addition, second-level analyses necessarily use relatively conservative thresholds for statistical significance, thus increasing the likelihood of a null result for a pre-defined location in the brain compared to testing that location in isolation. Thus, these six studies provide at best weak evidence against a direct neural effect of TMS on BOLD.

Importantly, the remaining six studies performed ROI-based analysis and thus provided much stronger evidence against a direct neural effect on TMS on BOLD activity. In the first three of these studies, the ROIs were defined for each subject based on either the activations from a different task (Baudewig et al., 2001) or anatomical considerations (Dowdle et al., 2018; Kearney-Ramos et al., 2018). The first study delivered supra-threshold stimulation (110% rMT) at 10 Hz for 1 second over the left lateral premotor cortex and found no activations near the presumed coil position in a second-level analysis (Baudewig et al., 2001). In addition, the authors also defined an ROI in the left lateral premotor cortex based on activations produced by a finger movement task. For that ROI, they examined the hemodynamic response function (HRF) produced by TMS and found that it does not differ from baseline. The second study (Dowdle et al., 2018) applied single pulses in a wide range of intensities (90-120% of rMT) over the left dorsolateral prefrontal cortex (DLPFC) and compared an active condition to a sham condition where the TMS coil was separated from the head by 3-cm wide foam. Finally, the third study was the largest concurrent TMS-fMRI study to date (N = 49). The authors delivered single-pulse TMS to right ventromedial PFC at 100% and compared the evoked BOLD activity to rest. Both of the last two studies reported no activation in the vicinity of the site of stimulation in second-level whole-brain analyses. In addition, both studies used pre-existing anatomical atlases to define anatomical ROIs and also found no significant BOLD activity in those ROIs. These findings thus serve as a counterweight to the positive findings by Bestmann et al., (2005) though they should also be interpreted with caution given the lack of specificity in the ROI definitions in all of these studies.

The strongest evidence to date, therefore, comes from the remaining three studies that reported on four different experiments and used the gold standard of precisely localizing the coil location for each subject (De Weijer et al., 2014; Kemna & Gembris, 2003; Rafiei et al., 2021). For each of these studies, the authors placed different types of markers directly on the TMS coil that were easy to localize in a T1-weighted (i.e., anatomical) MRI scan. Based on these markers, it was then possible to compute the precise location of the TMS coil on the head for each subject separately and thus define ROIs for the locations of maximum stimulation near the surface of the brain. We discuss each of these studies in more detail below.

In the first such study, supra-threshold TMS (150% rMT) was delivered to the left prefrontal and parietal cortex at 4 Hz for 1 second (Kemna & Gembris, 2003). The authors precisely determined the position of the TMS coil using a water-filled tube attached to it. They then defined an ROI for each subject at the site of stimulation and extracted the time-courses of BOLD activity following TMS stimulation. The results showed no significant correlation between the average BOLD time series and the canonical HRF in either the prefrontal or parietal cortex, even though a significant correlation was observed in a separate condition where TMS was delivered to M1. Further, whole-brain activation maps demonstrated that stimulation of prefrontal and parietal regions of the brain elicited activations in nearby areas, but the precise coil localization and ROI analysis demonstrated that the activations were not at the sites of stimulation. These results also demonstrate how easy it is to observe activation in the general vicinity of the TMS coil that is not in reality directly underneath it.

The second study to use the gold-standard localization method also employed water-filled tubes to precisely determine the location and orientation of the TMS coil (De Weijer et al., 2014). Importantly the study also measured the MRI signal quality at the exact spot of stimulation and did not find any signal dropout when compared to other brain regions. The authors compared supra-threshold (110% of rMT) with sub-threshold (70% of rMT) stimulation of the dorsal part of the supramarginal gyrus (dSMG) in three subjects (another six subjects received stimulation to M1 and these results were discussed in the previous section). Similar to the study by Kemna & Gembris, (2003), De Weijer and colleagues did not find significant activation at the precise site of stimulation.

The final study to deliver TMS outside of M1/V1 during rest and use the gold-standard localization employed a set of Vitamin E capsules positioned on the TMS coil and again localized the precise location of stimulation for each subject (Rafiei et al., 2021). The authors performed two experiments with a large number of TMS pulses, targeting DLPFC in both of them. In the first experiment, they included two conditions of supra-threshold stimulation (100% of rMT) and one condition of sub-threshold stimulation (50% of rMT). The first and third conditions delivered 10 TMS pulses at 1 Hz over 10 seconds, while the remaining supra-threshold condition delivered TMS in bursts of 4 pulses with 80 ms between individual pulses in a burst (i.e., at a rate of 12.5 Hz) with five bursts being given in a 10 second period (i.e., a rate of 0.5 Hz) for a total of 20 pulses. The authors defined several sets of ROIs of different sizes at the site of stimulation and found no difference between the supra- and sub-threshold conditions for any of them. In the second experiment, they used only 100% of rMT but delivered a substantially higher number of pulses: in four different conditions, 30 pulses were given over 1.2 seconds (25

Hz), 2.4 seconds (12.5 Hz), 3.6 seconds (8.33 Hz) or 6 seconds (5 Hz). Despite the very large number of pulses in each burst, no condition led to BOLD signal change in any of four different ROI sizes at the site of stimulation. Importantly, these null results occurred even though the TMS conditions could be decoded using multivoxel pattern analysis (MVPA) in the same ROIs in both experiments.

Summary

Overall, the evidence to date strongly suggests that TMS delivered at rest outside of the primary motor and visual cortices does not lead to increases in BOLD activity at the site of stimulation. This conclusion was reached in all four individual experiments that used the gold-standard technique of precisely localizing the actual TMS coil position by using markers placed directly on the TMS coil (De Weijer et al., 2014; Kemna & Gembris, 2003; Rafiei et al., 2021). Further, the conclusion is also supported by the majority of the published literature (13/16 studies), with the remaining three studies providing at best weak evidence for the notion that TMS leads to BOLD increases at the site of stimulation. Thus, the preponderance of evidence suggests that TMS delivered at rest outside of M1/V1 does not increase local BOLD activity.

**Studies that targeted the rest of the brain during task**

TMS may be expected to have different effects on BOLD depending on whether the targeted brain region is engaged in a task or not. All studies reviewed above employed TMS at rest with an emerging consensus that TMS has no direct effect on the BOLD activity at the site of stimulation outside of feedback from downstream areas for the case of M1 or V1 stimulation. However, it is unclear whether the same conclusions hold for the special case where the targeted brain area is already engaged in a task and thus may have elevated BOLD at the time of stimulation.

To date, nine studies have employed concurrent TMS-fMRI during a task (Bestmann, Swayne, et al., 2008; Bestmann et al., 2010; Feredoes, Heinen, Weiskopf, Ruff, & Driver, 2011; Heinen, Feredoes, Weiskopf, Ruff, & Driver, 2014; Heinen et al., 2011; Leitão, Thielscher, Tünnerhoff, & Noppeney, 2015; Nahas et al., 2001; Ricci et al., 2012; Sack et al., 2007). Of these, four reported TMS-induced BOLD activity at the site of stimulation, while five reported no such activity (**Table 4**). Here we examine these studies in more detail.

**Table 4. Studies conducted during a task**. The table lists all nine studies using concurrent TMS-fMRI and stimulating areas outside M1 or V1 during a task. The first four experiments reported significant activations in the vicinity of the TMS coil, whereas the remaining five reported no significant activations. aMT, active motor threshold; rMT, resting motor threshold; GLM, general linear model; HRF, hemodynamic response function; MSO, maximum stimulator output; PMd, dorsal premotor cortex; DLPFC, dorsolateral prefrontal cortex; FEF, frontal eye field; SPL, superior parietal lobule; PPC, posterior parietal cortex; IPS, intraparietal sulcus; R, right; L, left; ROI, region of interest.

| Study | Target | Task | Protocol | Contrast(s) | N | Activation | Analyses |
|---|---|---|---|---|---|---|---|
| Nahas et al. (2001) *Biol Psychiatry* | L DLPFC | Tone discrimination | 21 pulses over 21 s | 120% rMT > rest<br>120% rMT > 80% rMT | 5 | Yes | Second-level GLM |
| Bestmann et al. (2008) *Cereb Cortex* | L PMd | Grip | 5 pulses over 455 ms | 110% rMT > 70% aMT | 12 | Yes | 1) ROI (using coordinates from a previous study)<br>2) Second-level GLM |
| Feredoes et al. (2011) *PNAS* | R DLPFC | Working memory | 3 pulses over 270 ms | 110% rMT > 40% aMT | 16 | Yes | 1) ROI (using coordinates of the intended target) |
| Heinen et al. (2014) *Cereb Cortex* | R FEF | Visual attention task | 3 pulses over 270 ms | 110% rMT > 40% aMT | 16 | Yes | 1) ROI (using coordinates of the intended target) |
| Sack et al. (2007) *Cereb Cortex* | L SPL | Visuospatial judgment | 5 pulses over 300 ms | 100% MSO > task, no TMS | 8 | No | Second-level GLM |
| Bestmann et al. (2010) *J Neurosci* | Contralesional PMd | Grip | 5 pulses over 455 ms | 110% rMT > 70% aMT | 12 | No | Second-level GLM |
| Heinen et al. (2011) *Eur J Neurosci* | R angular gyrus | Visuospatial attention | 3 pulses over 270 ms | 120% rMT > 40% rMT | 5 | No | 1) ROI (using coordinates of the intended target)<br>2) Second-level GLM |
| Ricci et al. (2012) *Front Hum Neurosci* | R PPC | Line bisection task | single pulse | 115% rMT > no TMS | 3 | No | First-level GLM |
| Leitão et al. (2015) *J Neurosci* | R IPS | Spatial attention | 4 pulses over 400 ms | 69% MSO > sham TMS | 10 | No | Second-level GLM |

Studies reporting an increase in BOLD activity

The four studies that reported TMS-induced activity at the site of stimulation used a variety of different tasks and stimulation targets. The first study to employ concurrent TMS-fMRI during task targeted left DLPFC in blocks of 21 pulses delivered at 1 Hz while subjects engaged in a tone discrimination task (Nahas et al., 2001). The task was performed continuously in all conditions and thus it is impossible to determine whether it activated the stimulated region or not. The authors only performed a second-level analysis (no ROI analyses were conducted) and reported significant activations near the presumed location of the TMS coil for intensities of 120% of MT.

The remaining three studies improved on this first one by using predefined ROIs, though none of these studies defined the ROIs based on the actual coil position. The first study delivered five pulses of TMS at 11 Hz to PMd during a grip task or no grip rest (Bestmann, Swayne, et al., 2008). Supra-threshold (110% of rMT) stimulation was found to lead to BOLD increases compared to sub-threshold (70% of aMT) stimulation in a spherical ROI (15 mm diameter) defined based on previous research. The second study delivered 3 TMS pulses at 11 Hz to DLPFC time-locked to a working memory task known to activate DLPFC (Feredoes et al., 2011). Supra-threshold TMS (110% of rMT) led to higher BOLD than sub-threshold TMS (40% of rMT) in an ROI defined based on the targeted coordinates for each subject. The authors used neuronavigation outside the scanner to mark the point on the scalp but no measurement from inside the scanner was taken to determine how close the actual stimulation location was to the intended one. Finally, the third study also delivered 3 TMS pulses at 11 Hz but targeted the frontal eye field (FEF) during a time-locked attention task dependent on FEF (Heinen et al., 2014). The same stimulation intensity, procedure for defining the ROI, and ROI size were used as in the study by Feredoes et al., (2011). The results showed that supra-threshold TMS increased BOLD activity in the ROI compared to sub-threshold TMS in both of two attention conditions, but not in a passive viewing condition, suggesting that the effects of TMS were task-dependent.

Studies reporting no increase in BOLD activity

As a counterweight to the four studies that found TMS-induced increases of activity at the site of stimulation during tasks, five other studies failed to find such increases. Of these, four studies employed various attention tasks and targeted attention-related brain areas such as the superior parietal lobule (Sack et al., 2007), angular gyrus (Heinen et al., 2011), posterior parietal cortex (Ricci et al., 2012), and intraparietal sulcus (Leitão et al., 2015) in neurologically intact subjects. The remaining study employed a motor task and stimulated PMd (Bestmann et al., 2010) in a group of stroke patients. Three studies employed second-level analyses (Bestmann et al., 2010; Leitão et al., 2015; Sack et al., 2007), one employed first-level analyses (Ricci et al., 2012), and only one employed ROI analyses with the ROIs defined based on the targeted coordinates in a procedure equivalent to the one employed by Feredoes et al. (2011) and Heinen et al. (2014). Unfortunately, the lack of precise localization in the majority of these studies makes it hard to draw firm conclusions from them regarding the direct neural effects of TMS on the BOLD activity at the site of stimulation.

Summary

The concurrent TMS-fMRI studies that delivered TMS during a task thus far paint an inconsistent picture of whether BOLD increases under the coil are present. Four studies found such increases, whereas five did not. Critically, none of the studies used the gold-standard technique of localizing the site of stimulation for each subject based on the actual coil location, and five of the studies did not report ROI-based analyses at all. Further, while all four studies that reported a change in BOLD saw an increase, two of the studies that we have categorized as not showing changes in BOLD reported a decrease for at least one condition (Bestmann et al., 2010) or subject (Ricci et al., 2012). This heterogeneity in results makes it difficult to draw strong conclusions about either the presence or absence of TMS-induced effects at the site of stimulation during tasks. Nevertheless, at least one study suggests that it is indeed possible that the TMS effects at the site of stimulation are qualitatively different during task and rest (Heinen et al., 2014), though these results need to be replicated and extended before strong conclusions on the topic can be made.

**The effects of TMS on neuronal activity at the site of stimulation**

Our review of the literature suggests that TMS appears not to have any direct effect on the BOLD signal at the site of stimulation during rest. Instead, local BOLD activity changes are almost always due to feedback mechanisms (from induced finger movement or phosphenes), or, perhaps, caused by an interaction with BOLD activity increases brought about by accompanying tasks. In this section, we briefly review the TMS studies performed in non-human animals (hence referred to as “animals”) and discuss how what is known about the neuronal activity at the site of stimulation relates to the observed BOLD results.

To date, a total of seven studies have been published that recorded single neuron firing at the site of TMS stimulation in animals (Allen, Pasley, Duong, & Freeman, 2007; Kozyrev, Eysel, & Jancke, 2014; B. Li et al., 2017; Moliadze et al., 2003; Mueller et al., 2014; Pasley et al., 2009; Romero et al., 2019). The studies varied in many aspects including species (cats or monkeys), TMS delivery (single pulses or trains), state of the animals (anesthetized or awake), and location of stimulation (visual cortex, frontal eye field, parietal cortex, or motor cortex). More importantly, their findings are heterogeneous with no two studies from different labs finding similar effects of TMS. Nevertheless, one pattern of results appears in the majority of these studies: TMS typically produces a combination of both excitation and inhibition.

Three studies appear to suggest that TMS induces periods of increased and decreased firing that mostly cancel each other out. The first study delivered single TMS pulses to the primary visual cortex of anesthetized cats (Moliadze et al., 2003). The results suggested that TMS induces early excitation (up to 500 ms after TMS) and a weaker but long-lasting inhibition (up to 5-6 seconds after TMS). A more recent study delivered single TMS pulses to the caudal forelimb area (equivalent to primate motor cortex) of anesthetized rats (B. Li et al., 2017). The stimulation produced an early excitation (up to 50 ms after TMS), followed by a period of suppression (up to about 200 ms after TMS), and another period of excitation (up to 300 ms after TMS). No data were reported for the period after the initial 300 ms following stimulation. Finally, a third study delivered single TMS pulses to the visual cortex of anesthetized cats and used voltage-sensitive dye (VSD) imaging to monitor neural firing (Kozyrev et al., 2014). This study found a very brief excitation (< 20 ms after TMS) followed immediately by a prolonged period of suppression (up to 400 ms after TMS), which then turned into excitation again (lasting at least to 800 ms after TMS). No data were reported for the period after the initial 800 ms. Though this issue was never specifically examined, all three studies report results where the excitation and inhibition seem to mostly cancel each other out. Nevertheless, the fact that the last two studies only focused on a short period after stimulation (300 and 800 ms, respectively) makes it hard to determine whether this pattern of results would remain if the analyzed period were extended further. Nevertheless, what is clear is that in all three cases TMS produces alternating periods of excitation and inhibition. These results are also consistent with findings from an electrical stimulation study where stimulation resulted in a short excitatory response followed by a long-lasting inhibition (Logothetis et al., 2010).

Two more recent studies delivered single-pulse TMS to awake monkeys. The first study was primarily a technical report but it also showed that different neurons respond differently to TMS delivered over the frontal eye field with some neurons showing consistent increases in firing rate, while others showing consistent decreases (Mueller et al., 2014). However, the study did not report on the prevalence of each type of neuron, and it is thus unclear if TMS led to a systematic change in firing rate across the whole population of neurons at the site of stimulation. The second study targeted the parietal cortex and reported the existence of neurons that only showed increases in firing, neurons that only showed decreases, and neurons that showed periods of both increases and decreases (Romero et al., 2019) with most neurons showing increased firing. These results are again consistent with the notion that TMS induces a mixture of excitation and suppression.

Finally, two other studies delivered TMS to the visual cortex of anesthetized cats (Allen et al., 2007; Pasley et al., 2009). Both studies delivered bursts of stimulation ranging from 1 to 8 Hz and lasting between 1 and 4 seconds, and both studies recorded activity between 100 ms and 5 minutes after stimulation. Allen et al., (2007) found an increased activity that lasted for a full minute, but also a strongly suppressed response to visual stimulation for at least 5 minutes after TMS. Critically, a follow-up study by the same group showed that the increases observed in spontaneous activity diminish rapidly with the number of trials, with some neurons showing no overall change in firing already by the $5^{th}$ trial (Pasley et al., 2009). It should be noted that fMRI studies typically deliver hundreds of TMS trials, and therefore the BOLD effects in these studies primarily reflect the steady-state TMS influence on neural firing achieved after several dozen trials. In addition, Pasley et al., (2009) also showed that the neuronal response to TMS depends on the pre-TMS activity of the neurons, thus potentially explaining the inconsistent findings in fMRI studies conducted during different tasks.

In summary, the animal literature on TMS consistently finds that TMS produces complex effects on neuronal firing with most studies reporting a combination of both excitation and suppression. None of the studies to date establish whether TMS changes the total amount of firing over the first few seconds after TMS (i.e., the period relevant for the observed BOLD response) at the site of stimulation. Yet, at least some studies suggest that the TMS-induced excitation and inhibition may be balanced so well that there is little to no overall increase in firing during the first 1-2 seconds after TMS.

**Why does TMS not have any direct effect on local BOLD activity?**

The current review strongly suggests that TMS does not have a direct effect on local BOLD activity despite inducing a complex set of effects on the firing of neurons at the site of stimulation. Here we discuss several possible explanations for this finding.

One possible reason for the lack of BOLD increase at the site of stimulation is that there may be a signal drop underneath the TMS coil. Indeed, some of the studies that failed to find increased local BOLD activity attributed this lack of activation to low signal-to-noise ratio (SNR) in these areas (Blankenburg et al., 2008). However, while early concurrent TMS-fMRI setups had substantial current leakage that likely did affect SNR in the vicinity of the coil (Weiskopf et al., 2009), modern setups result in excellent signal underneath the TMS coil. This was, for example, demonstrated by a study that carefully examined SNR at the precise site of stimulation (based on gold-standard methods of in-scanner localization) and did not find any SNR decreases compared to other regions of the brain (De Weijer et al., 2014). Further, even if a decrease in SNR in the vicinity of the TMS coil was present in older studies, this decrease was never large enough to prevent the finding of significant BOLD activations in M1 when that region was targeted. Therefore, it does not appear that the lack of effect of TMS on the local BOLD activity outside of M1/V1 can be attributed to inadequate SNR.

Another possibility is that the BOLD response produced by TMS pulses does not follow the shape of the standard hemodynamic response function (HRF). In this view, TMS does lead to a significant BOLD response but this response cannot be detected in standard analyses that assume the HRF response. To test this possibility, several studies have performed finite impulse response (FIR) analyses to explore the actual shape of the BOLD response independently of any assumptions (Baudewig et al., 2001; Bestmann et al., 2003; Bohning et al., 1999; Bohning, Shastri, Wassermann, et al., 2000; Kemna & Gembris, 2003; Rafiei et al., 2021). These studies either found no significant BOLD increase at any time point (Baudewig et al., 2001; Bohning et al., 1999; Kemna & Gembris, 2003; Rafiei et al., 2021) or when a change was observed, such as after M1 stimulation, the response was similar to the standard HRF shape (Baudewig et al., 2001; Bestmann et al., 2003; Bohning, Shastri, Wassermann, et al., 2000; Kemna & Gembris, 2003). Thus, the lack of direct effect of TMS on local BOLD activity cannot be attributed to a non-standard response shape of the induced BOLD activity.

The explanation that we favor is that TMS does not produce a significant BOLD increase at the site of stimulation because it leads to a complex pattern of excitation and inhibition that mostly cancel each other out. This hypothesis is based on the several animal studies reviewed in the previous section that demonstrate that TMS induces periods of increased and decreased activity (Allen et al., 2007; Kozyrev et al., 2014; B. Li et al., 2017; Moliadze et al., 2003; Mueller et al., 2014; Pasley et al., 2009; Romero et al., 2019). Nevertheless, as already mentioned, no animal study to date convincingly demonstrates how TMS affects the total firing at the site of stimulation over 1-2 seconds after TMS, so this possibility remains speculative.

## Conclusion

Here we reviewed all published concurrent TMS-fMRI studies that report the effects of TMS on the BOLD activity at or near the site of stimulation. The review strongly suggests that TMS does not have any direct effect on local BOLD activity. The strongest evidence comes from studies that delivered TMS outside of M1/V1 during rest: the great majority of them find no BOLD changes at the site of stimulation. Studies targeting M1/V1 are also consistent with this interpretation since it appears that they only lead to BOLD increases in the presence of feedback related to hand twitches or phosphenes.

Throughout this review, we have repeatedly emphasized the need for precise localization of the actual location of stimulation via in-scanner measurements of the TMS coil position. We call this method the "gold standard" for localizing the TMS coil position and urge future studies exploring BOLD activity at the site of stimulation to always adopt it and report the results of ROI analyses.

We speculate that the most likely explanation for the lack of direct TMS-induced BOLD changes at the site of stimulation is that TMS induces periods of increased and decreased neuronal activity underneath the coil that balance each other out. This explanation suggests the presence of robust regulatory mechanisms that dynamically regulate the overall firing in an area in the presence of artificially induced firing. It is possible that such regulatory processes are disrupted in disorders such as epilepsy and that TMS could thus provide a promising avenue for studying their mechanisms.